\documentclass{svjour3}     
\smartqed  
\usepackage{graphicx}
\usepackage{colortbl}
\usepackage{xcolor}
\usepackage{ulem}
\usepackage{wasysym}
\usepackage{amssymb, latexsym, textcomp,  textcomp}
\begin{document}

\title{Observation of the Rydberg resonance in surface electrons on superfluid helium confined in a 4-$\mu$m deep channel}

\titlerunning{Rydberg resonance in surface electrons confined in a microchannel}        

\author {Shan Zou$^1$$^\dagger$, Sebastian Grossenbach$^2$, and Denis Konstantinov$^1$}


\institute{1. Quantum Dynamics Unit, Okinawa Institute of Science and Technology (OIST) Graduate University, Tancha 1919-1, Okinawa 904-0412, Japan.\\ 
2. \email{shan.zou@oist.jp}}

\institute{1. Okinawa Institute of Science and Technology, Tancha 1919-1, Okinawa 904-0412, Japan \\
2. Fachbereich Physik, Universit\"at Konstanz, DE-78457 Konstanz, Germany \\ 
     \email{shan.zou@oist.jp} }


\maketitle
\begin{abstract}

We report the first observation of the microwave-induced intersubband (Rydberg) resonance in the surface-bound electrons on superfluid helium confined in a single 4-$\mu$m deep channel. The resonance signal from a few thousand of surface electrons comprising the Wigner Solid (WS) is detected by observing WS melting due to the microwave absorption. The observed transition frequency for the two lowest Rydberg states, in the range of 0.4-0.5~THz, is determined by the image charges induced by the surface electrons in conducting electrodes of the microchannel and the potentials due to applied voltages, and is in a good agreement with our calculations. The observed large broadening of the resonance on the order of 10~GHz, which is due to inhomegeneous distribution of the pressing electric field in the microchannel, is also in a reasonable agreement with our finite-element modeling calculations. This study of the confined electrons is motivated by their potential use for charge and spin qubits.                     


\end{abstract}

\section{Introduction}
\label{intro}  

Electrons trapped on the surface of liquid helium provide us with an extremely clean and controllable charge system~\cite{Monarkha}. The surface-bound states of such electrons are formed due to their attraction to the induced image charges in the liquid and the hard-core repulsion from the helium atoms, which prevents the electrons to enter the liquid. The energy spectrum of such surface electrons (SE) consists of subbands of the quantized motion perpendicular to the surface and the free motion along the surface. The quantized Rydberg states of the perpendicular motion were proposed as an attractive system for qubit implementation, with a promise of sufficiently long coherence time of the qubit states~\cite{DykmanPRB2003}. However, a recent study shows that the relaxation time of the excited Rydberg states is shorter than 1~$\mu$s due to the inelastic ripplon scattering of electrons from the surface-capillary waves (ripplons)~\cite{ErikaPRL2021}. On the other hand, the spin states of surface electrons on liquid helium are expected to have longer coherence time (exceeding 100~s) than in any other solid-state materials, which makes them a viable resource for quantum computing~\cite{LyonPRB2004}. It was proposed that coupling of the spin states to the Rydberg states~\cite{ErikaPRL2019} or the states of confined lateral motion of an electron~\cite{SchuPRL2010} can greatly facilitate the spin-state manipulation and detection, which would open a new pathway towards building a scalable quantum computer. In the above proposals, the spin-orbit coupling is introduced by creating a difference in the Zeeman splitting of the orbital states in a sufficiently strong gradient of the applied magnetic field. Most recently, the electrons trapped on solid neon were recognized as another promising platform for realizing the spin qubits~\cite{JinQST2020}.    

Successful implementation of the proposed schemes for quantum-state detection and manipulation demands confinement and operation of electrons, at the level of a single particle, in a some kind of microstructured device. The circuit quantum electrodynamics (cQED) architecture comprising of a superconducting coplanar-waveguide (CPW) resonator integrated with an electron trap has already demonstrated capability for the single-electron quantum-state detection~\cite{SchuNat2019,JinArx2021}. The microchannel devices, which were extensively used to study the transport properties of SE on the superfluid helium~\cite{GlasPRL2001,IkegPRL2009,ReesPRL2016,BadrPRL2020}, might provide another very useful platform for such purposes. A typical device consists of an array of microchannels fabricated on a silicon substrate and filled with the superfluid helium by the capillary action. The surface of superfluid inside the channels is charged with SE, which can be shuttled along the channels by applying ac potentials to the conducting electrodes incorporated into the channel's structure. In addition, the applied electrostatic potential provides fine control of the number of electrons in the channels, for example realizing a one-dimensional (1D) chain of electrons along the channel~\cite{IkegPRL2012,ReesPRB2016}. Such devices have shown an unprecedented electron-transport efficiency by employing a charge-coupled device (CCD) configuration~\cite{BradPRL2011}, which is well suited for building a quantum-CCD architecture proposed for the trapped-ion quantum computing~\cite{WineNat2002}. Also, it was shown that the microchannel structure can be incorporated with other mesoscopic devices, such as a point-contact constriction, which enhance capabilities for charge manipulation~\cite{ReesPRL2011}.   

The above described microchannel devices might be also well suited for creation of a sufficiently large gradient of an applied magnetic field to realize the coupling between the spin states and orbital states of SE. For example, stripes of a ferromagnetic material, such as Cobalt or Permalloy, can be easily fabricated by lithography and evaporation, and placed along the channels. Magnetization of the material in an applied uniform magnetic field would produce a nonuniform stray field outside the material, while the close proximity of the stripes to SE confined in the microchannels ensures that the field gradient experienced by the electrons is sufficiently large. For example, our finite-element modeling (FEM) calculations show that 200-nm thick stripes of Cobalt placed at the bottom of a 4-$\mu$m deep channel filled with the superfluid helium would produce a gradient on the order $10$~Gauss/$\mu$m near the liquid surface. This would result in the difference between the Zeeman splittings for the ground state and the first excited Rydberg state of a surface-bound electron on the order of 1~MHz. This is comparable to the intrinsic linewidth of the corresponding Rydberg transition at temperatures below 1~K~\cite{LeaPRL2002}. 

Motivated by the possibility to realize the spin-orbit coupling of the spin states of SE and their Rydberg states, we performed the first measurement of the microwave-excited Rydberg resonance corresponding to the transition from the ground state to the first excited Rydberg state for SE confined in a single 4-$\mu$m deep, 20-$\mu$m wide and 100-$\mu$m long channel. Due to a small number of electrons, on the order of a few thousand, contained in a single channel the conventional methods of detection are not applicable. Therefore we employed a conductive detection of the Rydberg resonance by observing a resonance-induced change in the transport of electrons along the channel. Here we report the preliminary results of our study and discuss them in the light of further improvement of the detection methods towards developing the spin-state detection in SE.                                         			
					
\section{Methods}
\label{methods}

The conventional method to detect the intersubband resonance in a bulk SE sample containing about $10^8$ electrons is to observe the corresponding change in the transmitted power of the resonant microwave (MW) radiation due to its absorption by the excited electrons~\cite{LeaPRL2002,GrimPRL1974}. For the typical density of electrons $n_s\sim 10^8$~cm$^{-2}$, a single microchannel employed in our experiment would contain on the order of $10^3$ electrons, which makes the microwave absorption detection impossible. A new method of the Rydberg resonance detection was developed recently, which is based  on the measurement of the image-current induced by the excited SE in a conducting electrode capacitively coupled to the electrons~\cite{ErikaPRL2019}. While it was argued that this method can be potentially scaled down to the detection of the Rydberg excitation of a single electron, the currently achieved sensitivity of this method was not sufficient to detect the resonance in the experiment reported here. As an alternative, we used the conductive detection of the Rydberg resonance by observing a change in the current of SE driven along the microchannel, which is induced by the heating of electrons due to the resonant MW absorption~\cite{VolJETP1981,KonsPRL2007}. In particular, in order to increase sensitivity of the conductive detection we employed SE in a crystalline state to exploit a large change in the conductivity of SE upon their transition from the solid to the liquid phase.  

\begin{figure}[htt]
	\includegraphics[width=1.0\textwidth]{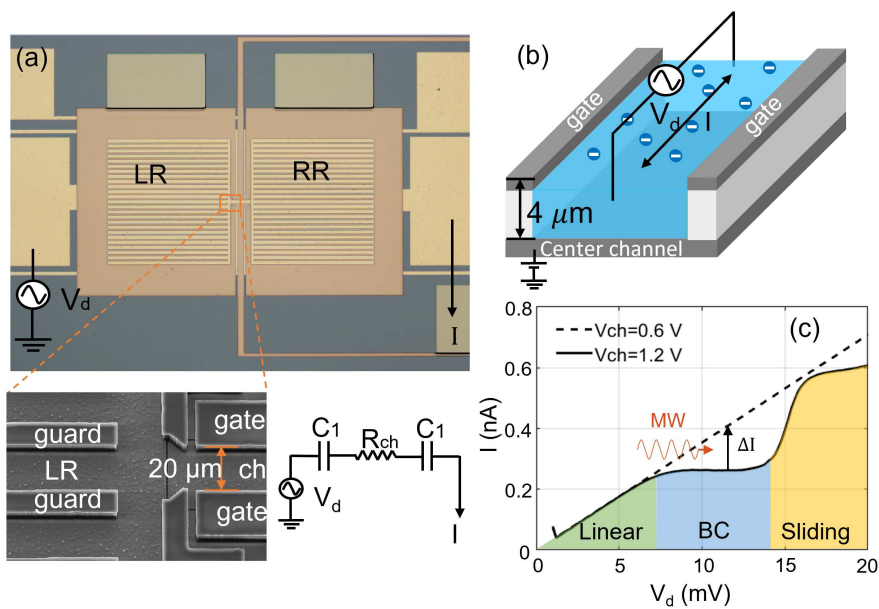}
	\caption{(color online) (a) Optical microscope image of the microchannel device. The driving ac voltage $V_\textrm{d}$ is applied to the bottom electrode of the left reservoir (LL) and is thereby coupled capacitively to the electrons on helium, while the current $I$ induced by the moving electrons is picked up at the bottom electrode of the right reservoir (RR). The inset shows a scanning electron micrograph of a portion of the 100-$\mu$m long central channel adjacent to the left reservoir. It also shows an equivalent lumped-circuit model of the device, as described in the text. (b) Schematics view of the central channel showing the electrode structure and simplified transport measurement scheme. (c) Measured I-V dependence of SE in the central channel indicating the different transport regimes, as described in the text. In particular, the dashed (solid) black line represents the I-V dependence measured at $T=180$~mK and $V_\textrm{ch}=0.6$~V (1.2~V) and corresponds to SE in the central channel in the liquid (solid) phase.}
	\label{fig:1}	 
\end{figure}

The microchannel device employed in our experiment is shown in Fig. \ref{fig:1}(a). The device consists of two identical arrays of 20-$\mu$m wide and 700-$\mu$m long microchannels connected in parallel, which serve as the left and right electron reservoirs (LR and RR, respectively). The two reservoirs are connected by a single 20-$\mu$m wide and 100-$\mu$m long central channel, which is depicted schematically in Fig. \ref{fig:1}(b). The whole structure is composed of two thin patterned gold layers separated by an insulating layer of hard-baked photoresist with a thickness of 4~$\mu$m, which defines the depth of the microchannels. The bottom gold layer consists of three electrodes which define the bottoms of the two reservoirs and the central channel. These electrodes are separated by 1~$\mu$m wide gaps, as can be seen in the inset of Fig.~\ref{fig:1}(a) which shows a magnified portion at the central channel adjacent to the left reservoir. The top gold layer consists of two electrodes, the split-gate and guard electrodes. The split-gate electrode, which is also shown schematically in Fig.~\ref{fig:1}(b), is aligned along the sides of the central channel. Together with the bottom electrode of the central channel, it serves to control the density of SE in the central channel by adjusting the electrostatic potential at the surface of liquid in the channel. Similarly, the guard electrode is aligned along the microchannels of each reservoir and serves to confine SE inside the microchannels. The corresponding electrical potentials applied to the electrodes of the bottom and top gold layers are denoted as $V_\textrm{LR}$, $V_\textrm{RR}$, $V_\textrm{ch}$, $V_\textrm{ga}$ and $V_\textrm{gu}$. 

The device was placed inside a leak-tight copper cell and cooled down at the mixing chamber of a dilution refrigerator. Liquid helium-4 was condensed into the cell, such that the liquid level settled slightly below the device, and filled the microchannels by capillary action. The liquid surface was charged with electrons produced by the thermionic emission from a tungsten filament placed about 1~mm above the device, with positive voltages $V_\textrm{LR}=V_\textrm{RR}=0.3$~V, $V_\textrm{ch}=1$~V applied to the reservoir and channel electrodes, while the gate and guard electrodes were grounded. After charging the surface, the reservoir and guard electrodes are typically maintained at these fixed values, while $V_\textrm{ch}$ and $V_\textrm{ga}$ are varied to change the number of electrons in the central channel. The charged surface of liquid in the microchannels, together with the gold electrodes, forming an equivalent electrical circuit is also shown in the inset of Fig.~\ref{fig:1}(a). Here, the capacitors $C_1$ and $C_2$ represent the effective capacitance between SE in the left and right reservoirs, respectively, and the electrodes of the device, while the resistor $R_\textrm{ch}$ represents the resistance of SE in the central channel~\cite{IkegPRL2009}. In the experiment, an ac voltage with the amplitude $V_\textrm{d}$ at the frequency of 20 to 50~kHz was applied to one of the reservoir's bottom electrodes, while the image-current $I$ induced by the SE motion was measured at the other reservoir's bottom electrode. The measured current could be fitted very well using the lumped-circuit model in the inset of Fig.~\ref{fig:1}(a), with the fitting parameter $C_1=C_2=0.6$~pF and the value of $R_\textrm{ch}$ defined by the conductivity of SE. Without MW excitation, the latter depends on the electron density $n_s$ in the central channel and the temperature of the liquid $T$.

There are two distinct regimes of the SE transport through the channel, both of which are illustrated in Fig.~\ref{fig:1}(c) showing the measured current $I$ versus the driving voltage $V_\textrm{d}$ for some typical experimental conditions. At a given temperature $T$ and low density $n_s$, SE are in the liquid phase. Correspondingly, the electron system exhibits a linear transport regime, with the electron current $I$ proportional to the driving voltage $V_\textrm{d}$, as shown by the dashed line in Fig.~\ref{fig:1}(c). At sufficiently large values of $n_s$, such that $\Gamma \gtrsim 137$, where $\Gamma=e^2(\pi n_s)^{1/2}/(4\pi\epsilon_0 k_B T)$ is the plasma parameter ($\epsilon_0=8.85\times 10^{-12}$~F/m is the vacuum permittivity), SE crystallize into the Wigner solid (WS) phase with the hexagonal lattice structure~\cite{Monarkha}. In the solid phase, the transport of SE is strongly nonlinear, as illustrated by the solid line in Fig.~\ref{fig:1}(c). In particular, three different regimes of the WS transport can be distinguished in the experiment, as indicated by the color regions in  Fig.~\ref{fig:1}(c). For sufficiently small values of the driving voltage $V_\textrm{d}$, WS exhibits the linear transport, similar to SE on the liquid phase. As the driving voltage increases, the measured current strongly deviates from the linear dependence on the driving voltage and saturates at a constant value (about 0.26~nA in Fig.~\ref{fig:1}(c)) at sufficiently large values of $V_\textrm{d}$. This nonlinear transport feature is due to the coherent Bragg-Cherenkov (BC) emission of resonant ripplons on the surface of the liquid by the driven WS. The WS lattice causes a commensurate deformation of the liquid surface. As the lattice is driven along the surface, it emits ripplons whose wave vector coincides with the reciprocal-lattice vector of WS. The amplitude of such ripplons, therefore the commensurate deformation of the liquid surface, is resonantly enhanced due to the constructive Bragg interference, which in turn leads to the enhancement of the effective mass of SE due to their coupling to ripplons~\cite{DykmPRL1997}. This leads to the decrease in the electron conductivity, therefore the measured current $I$ does not increase further as $V_\textrm{d}$ increases. When the driving voltage $V_\textrm{d}$ exceeds the critical voltage, WS eventually decouples from the surface deformation, which leads to a sudden increase in the conductivity and measured current. This is the sliding regime of the WS transport~\cite{ShirPRL1995}.                         

The nonlinear transport features of WS described above suggest an interesting possibility to detect the microwave-induced Rydberg resonance of SE confined in the central channel. It is well known that the resonant MW excitation of the Rydberg transition can result in a strong overheating of the electron system due to the microwave absorption and elastic decay of the excited electrons~\cite{KonsPRL2007}. Suppose that at a given driving voltage $V_\textrm{d}$ the current of WS in the central channel is set at the BC plateau, see Fig.~\ref{fig:1}(c). The resonant excitation of SE will cause the heating of SE and, at sufficiently large MW power, melting of WS. As a result, the observed current must experience an abrupt increase by $\Delta I$ to its value corresponding to the linear transport of SE in the liquid phase, as indicated in Fig.~\ref{fig:1}(c). Alternatively, even if the applied MW power is not sufficient to melt WS, the heating will most likely induce disorder in the electron crystal lattice, which would facilitate the WS sliding and enhancement in the measured current $I$. This method of the Rydberg resonance detection is used in the experiment reported here.                  

\section{Experimental results}
\label{results}

In the experiment, the microwave radiation at a tunable frequency in the range $\omega/2\pi=400$-500~GHz is introduced into the cell containing the microchannel device from a room-temperature source via a MW guiding system. The main part of the guiding system is an overmoded (WR-28) stainless-steel waveguide, while its short portion entering the cell is an overmoded (WR-06) copper waveguide sealed with a piece of 50-$\mu$m thick Kapton film. The input MW power could be adjusted using an attenuator at the output of the MW source. At a given attenuator setting, the power output of our MW source strongly depends on the MW frequency. Therefore, in the experiment we kept the frequency of MW radiation fixed and tuned the Rydberg transition frequency $\omega_{21}$ of SE in the central channel via the Stark effect by adjusting the dc (pressing) electric field $E_\perp$ acting on SE perpendicular to the surface of liquid~\cite{LeaPRL2002}. Practically, this was done by varying the voltage $V_\textrm{ch}$ applied to the bottom electrode of the central channel. We note that simultaneously it caused variation of the electron density $n_s$ in the central channel, which increases with increasing $V_\textrm{ch}$.         

Due to the upper-frequency limit of our MW source, the typical values of $V_\textrm{ch}$ required to tune SE into resonance ($\omega_{21}=\omega$) were significantly lower than $1.2$~V, the value used to obtain the nonlinear transport curve shown in Fig.~\ref{fig:1}(c) (the solid black line). The useful range of $V_\textrm{ch}$ for the given bandwidth of our source was found to be in the range 0.7-0.9~V. Figure 2(a) shows the I-V curves of SE in our device taken without MW excitation for a fixed gate voltage $V_\textrm{ga}=0$ and several different values of $V_\textrm{ch}$ in the above range. At $V_\textrm{ch}=0.6$~V the electron system shows the linear I-V dependence corresponding to SE in the liquid phase. With increasing $V_\textrm{ch}$, therefore the increasing density $n_s$, the I-V curves start exhibiting nonlinear features, in particular a series of plateaus followed by the abrupt rises in current. This behavior was recently identified with the repetitive coupling and decoupling of WS to and from the liquid surface deformations~\cite{ZouPRB2021}, which is sort of an intermediate regime between the linear transport and strong pinning of WS in the resonant BC-emission regime. The observed nonlinear transport of WS still makes it suitable for the proposed detection of the Rydberg resonance by the electron heating which affects WS and alternates its transport properties, although with a reduced sensitivity compared  with the case discussed in relation to Fig.~\ref{fig:1}(c).       

\begin{figure}[htt]
\centering
	\includegraphics[width=0.8\textwidth]{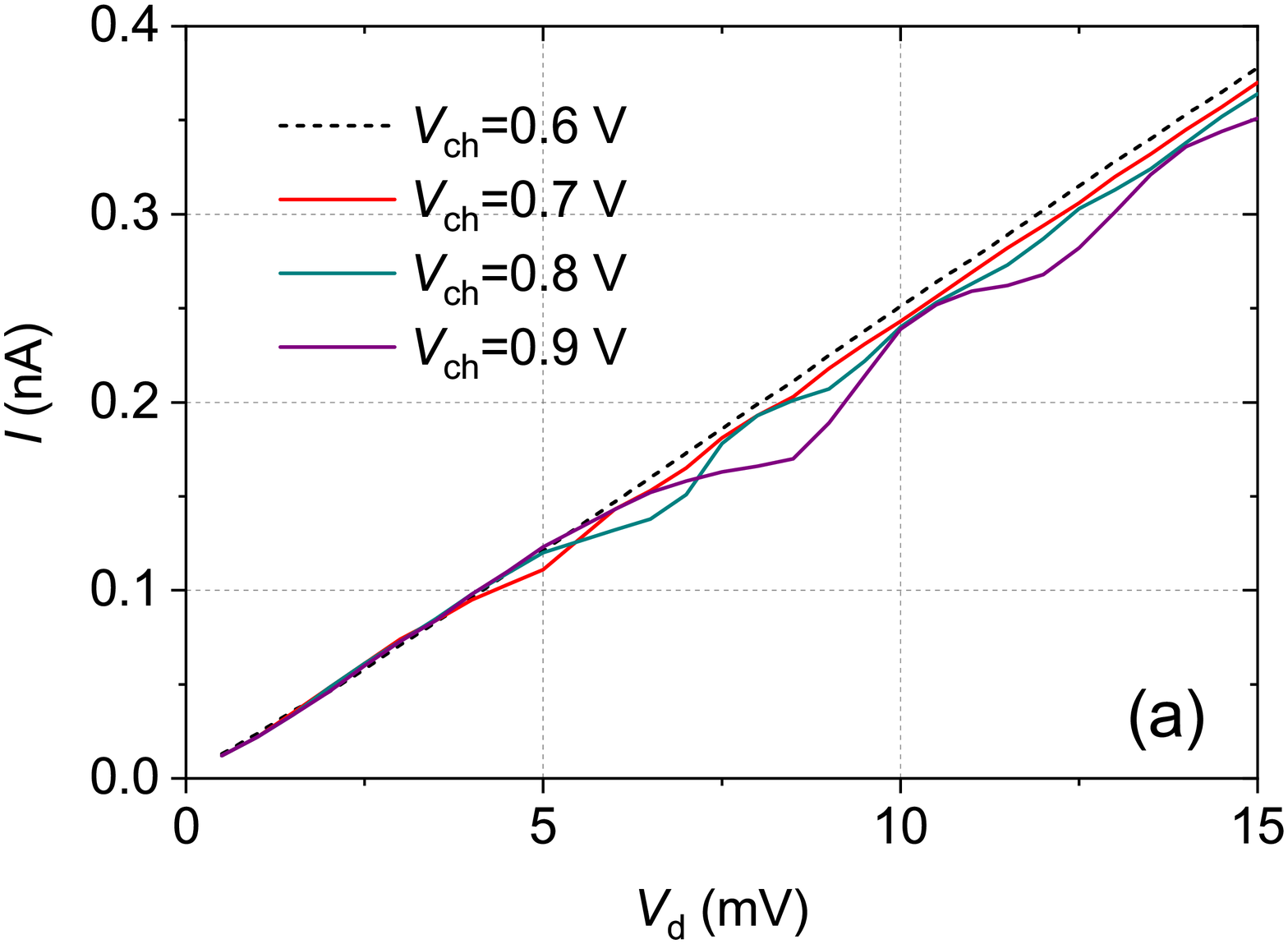}
	\includegraphics[width=0.8\textwidth]{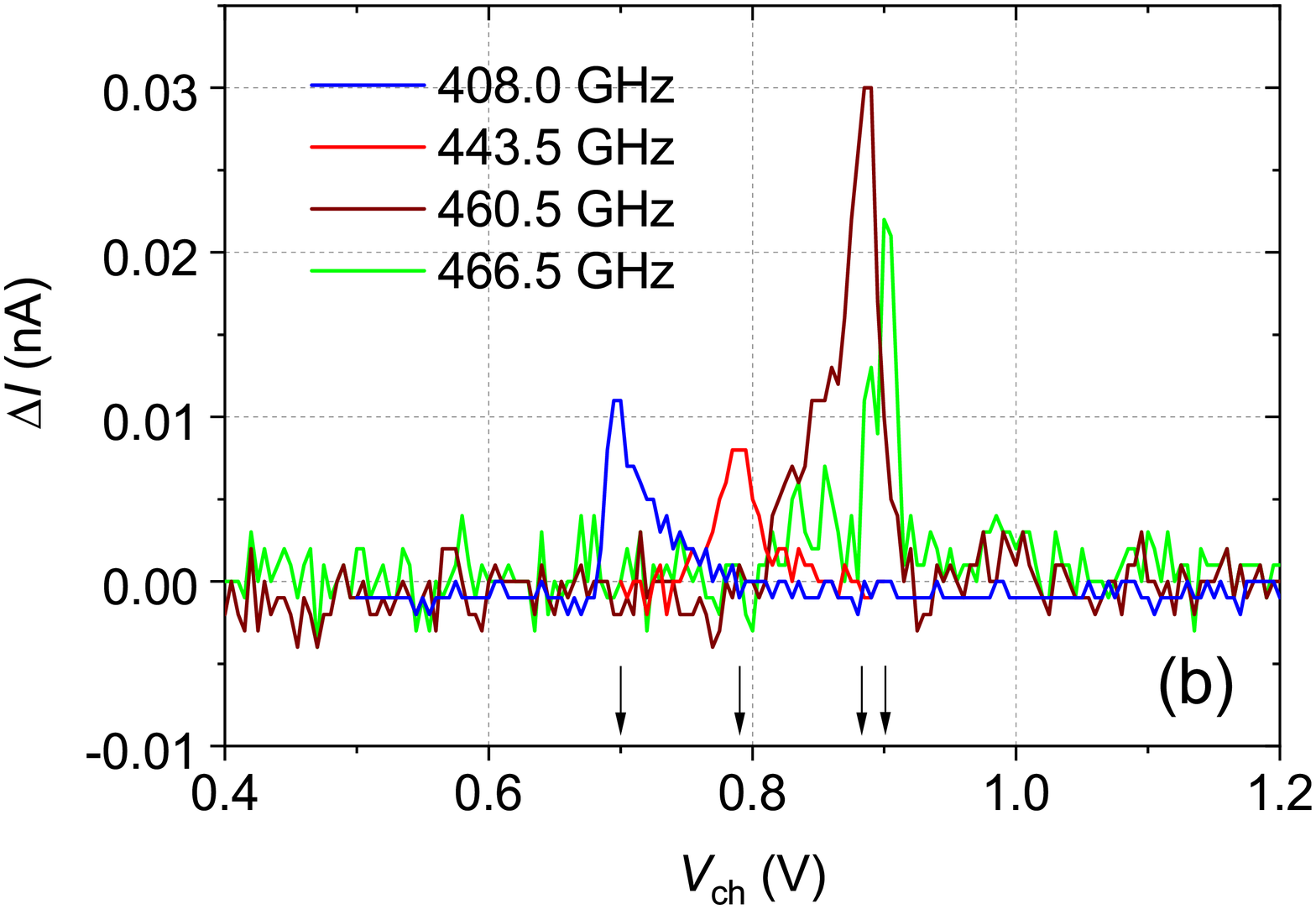}
	\caption{(color online) (a) I-V dependence of SE in the central channel measured without MW radiation at $T=150$~mK for different values of the channel voltage $V_\textrm{ch}$ and fixed gate voltage $V_\textrm{ga}=0$. (b) Variation of the change in current magnitude $\Delta I$ with the channel voltage for SE under irradiation for several values of the MW frequency. The values of $\Delta I$ were obtained from the measured current by subtracting the values of current measured without the radiation. An abrupt change in current indicates the resonant MW-induced transitions of SE from the ground state to the first excited Rydberg state. Arrows indicate resonant values of $V_\textrm{ch}=0.7$, 0.78, 0.88 and 0.9~V corresponding to the peak values of $\Delta I$ for each MW frequency.}
	\label{fig:2}	
\end{figure}

Figure~\ref{fig:2}(b) shows variation of the change in current magnitude $\Delta I$ with the channel voltage $V_\textrm{ch}$ and fixed gate voltage $V_\textrm{ga}=0$ for SE exposed to radiation at different values of MW frequency. In order to obtain the change in current $\Delta I$ upon irradiation, the current data obtained without radiation were subtracted. The current was measured for a range of driving voltages $V_\textrm{d}$ from 2.5 to 6.5 mV. Due to slight overheating of the experimental cell by the MW radiation, the temperature $T$ varied in a range from 170 to 190 mK for different traces. An abrupt change in current $I$ with respect to its value without radiation indicates overheating of the electron system by the microwave absorption which accompanies resonant transitions between the Rydberg states. As shown in the following section, this corresponds to the transition of SE in the central channel from the ground state to the first excited Rydberg state. As expected, the position of the resonance shifts towards the higher values of $V_\textrm{ch}$ with increasing MW frequency due to the Stark effect. 

The measured resonance signal exhibits large broadening and a rather complicated line shape. The observed signal broadening, on the order of 10~GHz, is many orders of magnitude larger than the intrinsic linewidth of the Rydberg transition~\cite{LeaPRL2002}. As shown in the following section, this large broadening arises due to the inhomogeneous broadening of the Rydberg transition of SE in the central channel because of the variation of the pressing electric field $E_\perp$ across the channel. It is harder to account for the rather irregular asymmetric shape of the signal.  In addition, it should be noted that the detection method realized here does not measure the transition rate directly, instead, it probes the electron conductivity in response to the transition-induced heating of SE. In order to fully account for the shape of such a conductivity response one has to consider dependence of the electron temperature on $V_\textrm{ch}$, the corresponding change in the conductivity and measured current, etc. Also note that the electron density $n_s$ in the central channel varies with $V_\textrm{ch}$. Accounting for all these effects is a rather formidable task at the moment and is beyond the scope of the present report.     

\section{Discussion}
\label{discussion}

The pressing electric field $E_\perp$ acting on SE in the central channel is due to the charges induced in the channel and gate electrodes by SE and voltages applied to the electrodes. An approximate relationship between the values of $E_\perp$ and $V_\textrm{ch}$ can be established by using a simplified parallel-plate capacitor model, where we represent the surface of liquid helium in the central channel charged with SE by a layer of negative charge with a uniform density $-en_s$ ($e>0$ is the elementary charge) and a fixed potential $V_e$. Together with the channel bottom electrode at the potential $V_\textrm{ch}$, such a layer forms a parallel-plate capacitor with the distance between the plates $d$ equal to the height of the central channel walls. The pressing electric field acting on SE perpendicular to the surface of liquid is given by $E_\perp=(V_\textrm{ch}-V_e)/d-en_s/(2\epsilon_0)$. Using the same model and taking into account coupling of SE to the gate electrode of the central channel, we can express the electron density as~\cite{LinJLTP2019}

\begin{equation}
n_s=\frac{\epsilon \epsilon_0}{\alpha e d} \left( \alpha V_\textrm{ch} + \beta V_\textrm{ga} - V_e \right), 
\label{eq:ns}
\end{equation}

\noindent where  $\alpha$ and $\beta$ are the weighted contributions ($\alpha + \beta = 1$) to the total capacitance between SE and the channel and gate electrodes, respectively, and $\epsilon$ is the dielectric constant of liquid helium which very close to 1. For $V_\textrm{ga}=0$ corresponding to the condition under which the data shown in Fig.~\ref{fig:2} where taken, we obtain an approximate analytical relationship between $E_\perp$ and $V_\textrm{ch}$

\begin{equation}
E_\perp = \frac{V_\textrm{ch}}{2d} - \frac{V_e}{d} \left( 1-\frac{1}{2\alpha} \right). 
\label{eq:Eperp}
\end{equation}

\noindent Note that the potential $V_e$ represents the potential of SE in the reservoirs and can be determined experimentally from the values of the channel and gate voltages at which the measured current $I$ becomes zero, which corresponds to the vanishing density of SE in the central channel, see Eq.~(\ref{eq:ns}). From the same procedure, the value of $\alpha$ can be determined~\cite{LinJLTP2019}.

\begin{figure}[htt]
\centering
	\includegraphics[width=0.82\textwidth]{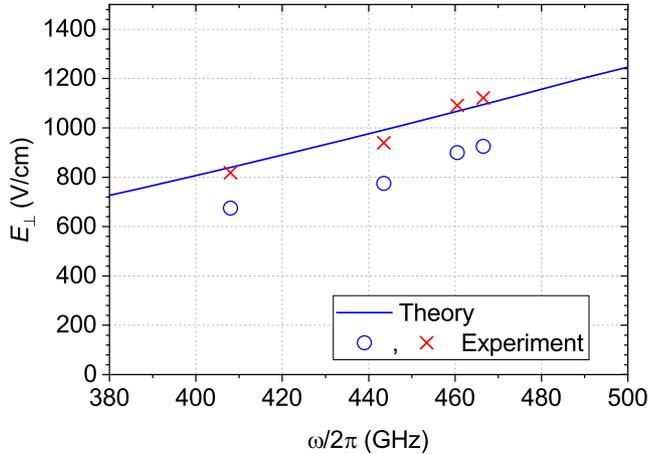}
	\caption{(color online) Pressing electric field $E_\perp$ corresponding to the resonant MW frequency $\omega$ for the transition from the ground state to the first excited Rydberg state calculated using the infinite-barrier approximation (solid line). The blue open circles and red crosses represent the pressing field determined from the experimental data shown in Fig.~\ref{fig:2}(b) using the parallel-plate capacitor model with $d =4$ and $3.3$~$\mu$m, respectively, as described in the text.}
	\label{fig:3}	
\end{figure}

Figure~\ref{fig:3} shows the values of the pressing field $E_\perp$ corresponding to the peak position of the resonance signals shown in Fig.~\ref{fig:2}(b) for each microwave frequency $\omega$ (blue open circles). These values are obtained from the corresponding values of $V_\textrm{ch}$ indicated by arrows in Fig.~\ref{fig:2}(b) using Eq.~(\ref{eq:Eperp}) with $d=4$~$\mu$m, the height of the central channel walls, and $V_e=0.2$~V, $\alpha=0.8$ determined from the experimental data. For the sake of comparison, the solid line shows the relationship between the frequency $\omega_{21}$ and $E_\perp$ calculated by solving the 1D stationary Schrodinger equation for the electron motion perpendicular to the liquid surface assuming an infinitely large potential barrier at the liquid-vacuum interface~\cite{Monarkha}. The experimental data exhibits a similar slope of the frequency-field dependence, which indicates that the observed resonance corresponds to the MW-excited transitions from the ground state to the first excited Rydberg state. At the same time, the values of $E_\perp$ estimated from the experimental data are noticeably  lower than the theoretical ones. The most probable cause for such discrepancy is the parallel-plate capacitance approximation, where we assume a flat surface of liquid helium with the depth $d$ equal to the height of the channel walls. The actual surface of liquid in a microchannel is curved due to the capillary effect and pressure exerted by SE on the surface of liquid helium~\cite{KliePRB2002}. A better approximation is a surface having the round shape with the radius of curvature given by $R_c=\sigma/\left( \rho g h +n_seE_\perp \right)$, where $\sigma$ and $\rho$ is the surface tension and density of liquid helium, respectively, $g$ is the acceleration due to gravity, and $h$ is the height of the microchannel device above the bulk liquid level. We note that the exact value of $h$ can not be measured in our experiment, therefore we did not attempt to take into account the deviation of the surface shape from the flat one. Nevertheless, the depression of the surface of liquid due to the curvature can be roughly taken into account by assuming somewhat lower depth of the liquid $d$ compared with the height of the channel walls. For example, the red crosses in Fig.~\ref{fig:3} show the values of $E_\perp$ estimated from the experimental data using Eq.~(\ref{eq:Eperp}) assuming $d=3.3$~$\mu$m. Thus, we conclude that the approximate parallel-plate model already accounts reasonably well for the obtained experimental data.  

\begin{figure}[htt]
\centering
	\includegraphics[width=0.78\textwidth]{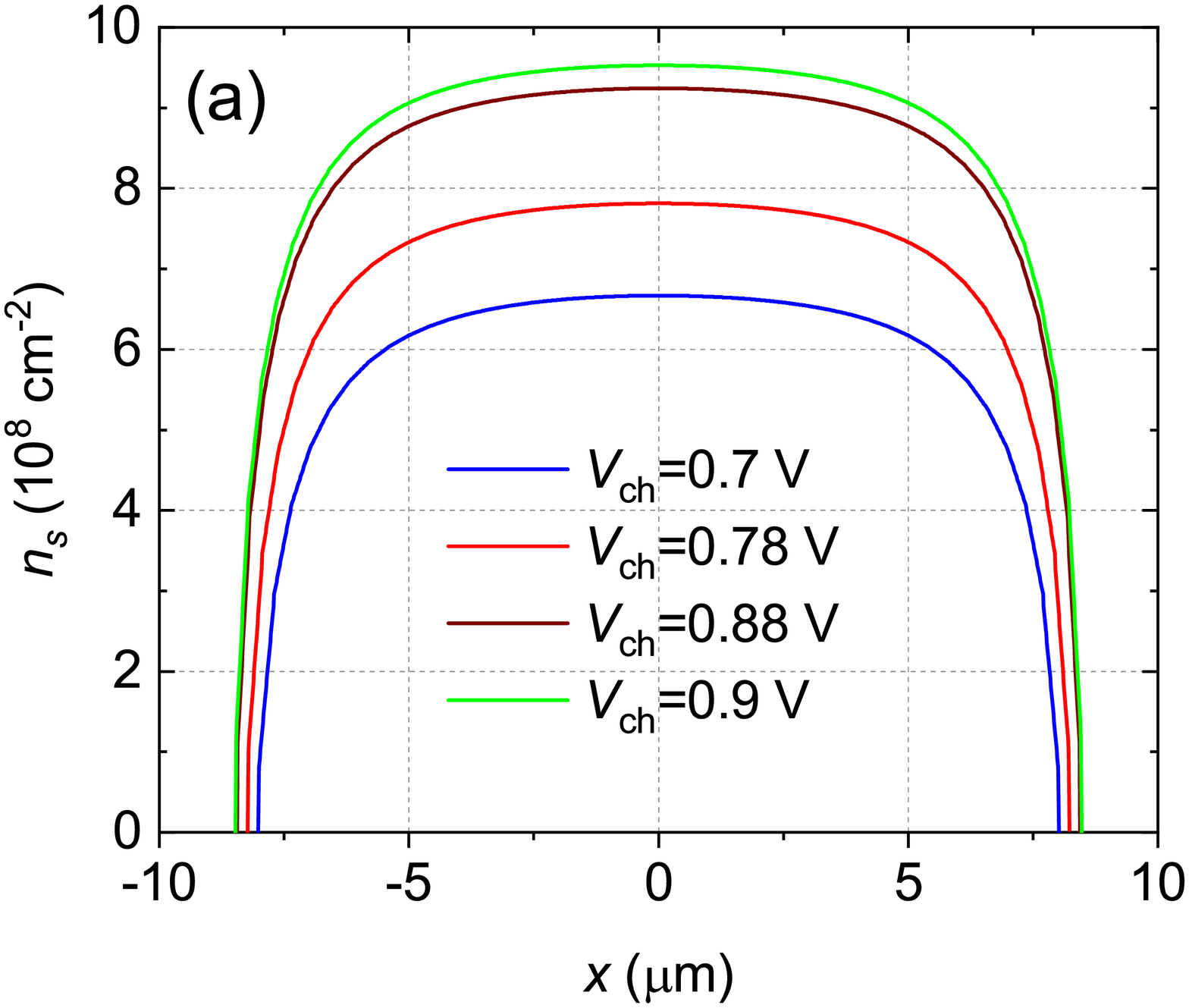}
	\includegraphics[width=0.78\textwidth]{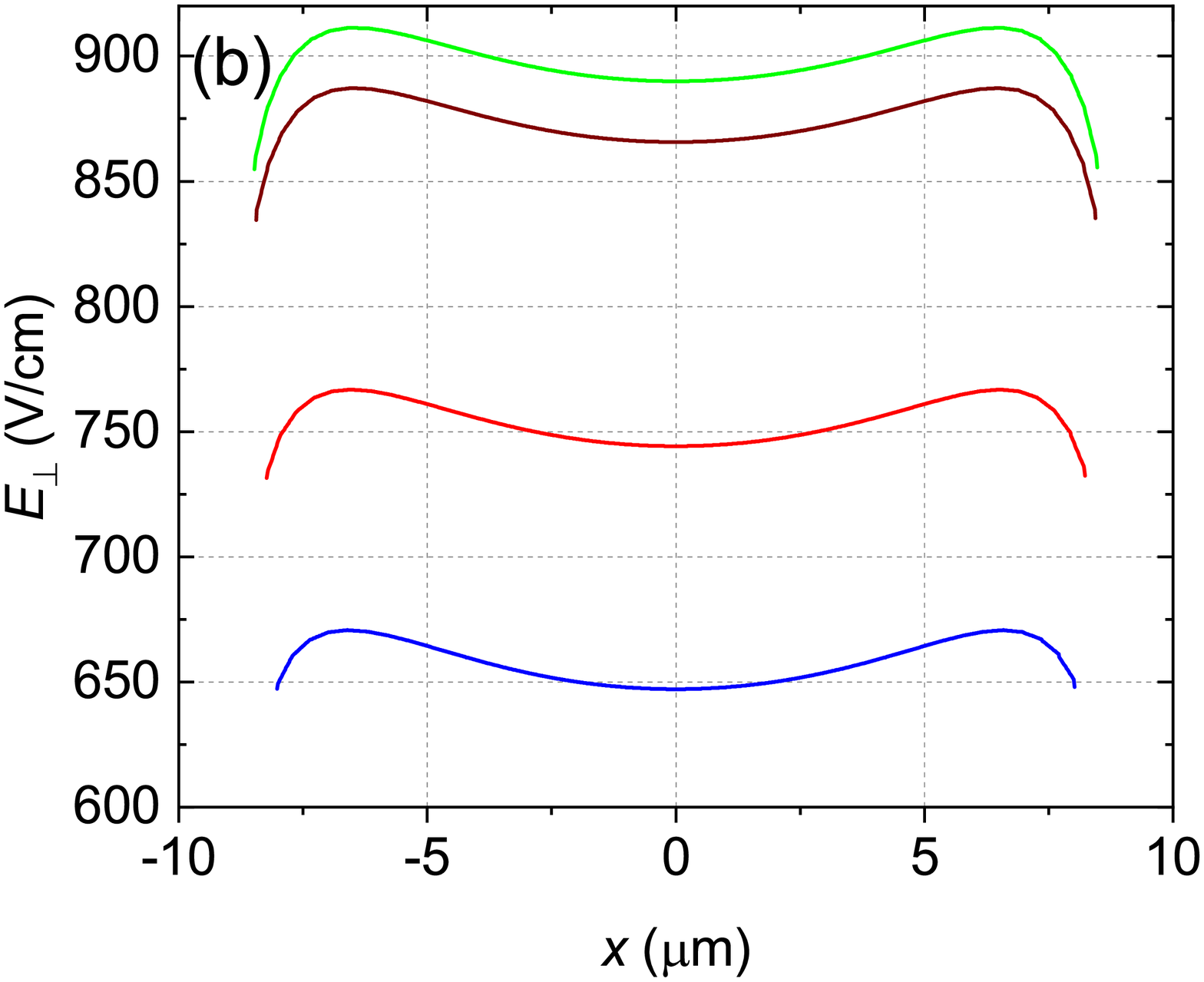}
	\caption{(color online) (a) The electron density $n_s$ in the central channel versus the distance from the center of the channel calculated for different values of the channel voltage $V_\textrm{ch}$ and fixed gate voltage $V_\textrm{ga}=0$. In the simulation, the SE are represented by an equipotential plane at potential $V_e=0.2$~V, as described in the text. (b) The corresponding cross-sectional profile of the pressing electric field $E_\perp$ acting on the electrons.}
	\label{fig:4}	
\end{figure}

The above model does not provide estimations for the distribution of $E_\perp$ across the electron system, which is important to know in order to account for the broadening of the Rydberg resonance line, as discussed below. In order to get more insight, we performed FEM calculations of the electric potential distribution across the channel for an infinitely long channel using the COMSOL Multiphysics simulation software. Similar to the parallel-plate capacitor model, in our numerical simulations we assume a flat surface of liquid helium in the channel and model SE as an equipotential plane at potential $V_e$. For a given value of $V_e$, the equilibrium width of the charged surface in the channel is found by setting the in-plane electric field at the edge of the charge system to zero. Then, the electron density profile across the channel is found from the difference of the perpendicular electric field below and above the charged helium surface (as before, we assume $\epsilon \approx 1$), using the Gauss's law, while the pressing electric field $E_\perp$ acting on SE is given by the arithmetic mean of the above two fields. Figure~\ref{fig:4}(a) shows the equilibrium density profiles across the channel calculated for several values of $V_\textrm{ch}$ (these values are indicated by the arrows in Fig.~\ref{fig:2}(b)). This shows that for the typical values of $V_\textrm{ch}$ used in our experiment the width of the electron system across the channel is close to the width of the channel. Figure~\ref{fig:4}(b) shows the corresponding distribution of the pressing electric field $E_\perp$ across the electron system. In this simulation, we assumed the depth of the liquid in the channel $d=4$~$\mu$m, which coincides with the height of the channels walls. Note that the average values of $E_\perp$ across SE for each channel voltage are fairly close to those obtained from the parallel-plate capacitor model, see Fig.~\ref{fig:3} (open circles). 

As seen in Fig.~\ref{fig:4}(b), the simulations reveal the variation of $E_\perp$ on the order of 30~V/cm across the electron system. This variation leads to the inhomogeneous broadening of the Rydberg transition line for the many-electron system. For a given fixed value of the MW frequency, a portion of SE in the channel experiencing the pressing electric field with a certain resonant value $E_\perp^{(r)}$ will be tuned in resonance with the radiation. The location of this portion and the corresponding number of resonant SE will depend on the channel voltage $V_\textrm{ch}$. As $V_\textrm{ch}$ is increased from a value far below the resonance, it is reasonable to expect that SE away from the center of the channel, which experience the maximum $E_\perp^{(\textrm{max})}$ at a given channel voltage, will be resonantly excited by the radiation, when $E_\perp^{(\textrm{max})}=E_\perp^{(r)}$. As $V_\textrm{ch}$ increases further, these electrons will become detuned from the resonance, while the electrons closer to the middle of the channel will become resonantly tuned. As mentioned earlier, it is difficult to account for the exact shape of the resonant signal as SE are swept through the resonance by the varying field $E_\perp$. Nevertheless, it is reasonable to expect that the broadening of the observed resonance lines will be determined by the variation of $E_\perp$ across the system. Note that the variation of $E_\perp$ observed in the simulations corresponds to the variation of the transition frequency $\omega_{21}$ on the order 10~GHz, as determined from the slope of the theoretical line in Fig.~\ref{fig:3}. This is in a reasonable agreement with the broadening of the measured resonance signals shown in Fig.~\ref{fig:2}(b).     
  
\section{Conclusion}
\label{Conclusion}

We presented the results of an experiment where the microwave-induced transitions between the ground and first-excited Rydberg states are observed for the first time for the surface electrons on superfluid helium confined in a micron-size channel with dimensions much less than the radiation wavelength. To observe the resonance signal from a few thousand of electrons, we employed a conductive-detection method by observing a change in the electron current through the channel in response to the resonant heating of SE by the microwave absorption. The observed large broadening of the resonance line on the order of 10~GHz is due to the inhomogeneous distribution of the pressing electric field acting on SE and is in a reasonable agreement with our numerical calculations. It seems to be feasible to significantly reduce this broadening by decreasing the width of the electron system in the channel. In particular, it has been already demonstrated that such width can be reduced to effectively a single electron across the channel~\cite{ReesPRB2016,LinJLTP2019}. At the same time, the reduction of the total number of SE in the channel would require development of an ultra-sensitive detection method. Such an ultra-sensitive method can be based on the image-charge detection demonstrated earlier, providing further improvements in the detection circuit are developed~\cite{ErikaPRL2019}. The experiment described here was motivated by the potential realization of the spin-orbit coupling in SE and detection of the spin states of electrons, thus can be considered as a first step towards the spin qubits on liquid helium.

%

\begin{acknowledgements}
The work was supported by an internal grant from the Okinawa Institute of Science and Technology (OIST) Graduate University and Grant-in-Aid for Scientific Research (Grant No. 17H01145) KAKENHI MEXT. S. G. acknowledges support of an OIST Graduate University internship program.
\newline

\noindent \textbf{Data Availability Statement} The datasets generated during the current study are available from the authors on reasonable request. 
\end{acknowledgements}




\end{document}